\documentclass[12pt,letterpaper]{article}
\usepackage{jheppub}
\usepackage{amsmath, amssymb}
\usepackage{verbatim}
\usepackage{graphicx}
\usepackage{subfigure}
\usepackage{epsfig}
\usepackage{epstopdf}
\usepackage{amsthm}
\usepackage{xcolor}
\usepackage{esint}
\input{epsf}

\newcommand{\rbr}[1]{\left(#1\right)}

\title{On Intersecting Conformal Defects}

\author{Tom Shachar}
\affiliation{The Racah Institute of Physics, The Hebrew University of Jerusalem, \\ Jerusalem 91904, Israel}
\emailAdd{tom.shachar@mail.huji.ac.il}

\abstract{
	We study the physics of 2 and 3 mutually intersecting conformal defects forming wedges and corners in general dimension. For 2 defects we derive the beta function of the edge interactions for infinite and semi-infinite wedges and study them in the tricritical model in $d=3-\epsilon$ as an example. We discuss the dependency of the edge anomalous dimension on the intersection angle, connecting to an old issue known in the literature. Additionally, we study trihedral corners formed by 3 planes and compute the corner anomalous dimension, which can be considered as a higher-dimensional analog of the cusp anomalous dimension. We also study 3-line corners related to the three-body potential of point-like impurities.
}

\begin{document}
	\maketitle
	
\section{Introduction}
Defects in QFT and lattice systems is a vast and rapidly growing field. A particularly well-studied kind are defects that are created by altering the interactions on some lower-dimensional surface inside the bulk. From the CFT perspective, they can be thought of as conformal theories locally deformed on some sub-manifold. Many interesting universality classes can be explored in this way, usually by deforming a bulk CFT or BCFT/DCFT with conformal boundary conditions such as Dirichlet or Neumann. This approach enjoys access to powerful tools such as Bootstrap, epsilon expansion, and large N methods \cite{Billo2016,PhysRevB.104.104201,SciPostPhys.15.3.090,Beccaria2022,Giombi2023,Giombi2023a,10.21468/SciPostPhys.15.6.240,Shachar2024,Trepanier2023a,RavivMoshe2023a,PhysRevLett.128.021603,Cuomo2022,PhysRevLett.130.151601,Aharony2023,Cuomo2024,Herzog:2020lel,Lauria2021,Herzog2022,Gimenez-Grau:2022ebb,Bianchi:2022sbz,Bianchi2024,Dey:2024ilw,RodriguezGomez2022,RodriguezGomez2022a,CarrenoBolla2023,Nishioka2023,Harribey2023,Harribey2024gjn,PhysRevD.96.045003,Drukker2020,PhysRevLett.129.201603,Billo2023,Billo2024,Dempsey:2024vkf,Karateev2024skm,Pannell2023,Nagar2024,Pannell:2024hbu,Bashmakov:2024suh,Zhou:2024dbt,Hu2024,Diatlyk:2024ngd,Bianchi:2024eqm}.

\begin{figure}[h!]
	\centering
	\includegraphics[width=0.25\columnwidth,trim={11cm 4cm 12cm 5cm},clip]{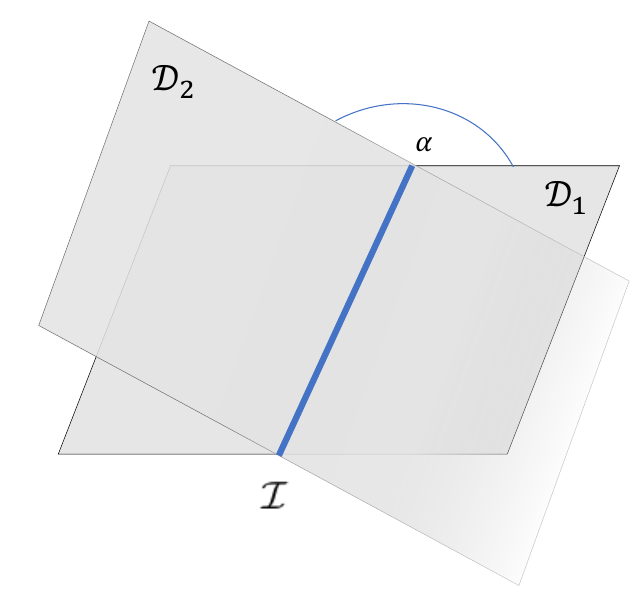}\quad
    \includegraphics[width=0.25\columnwidth,trim={12cm 5cm 13cm 5cm},clip]{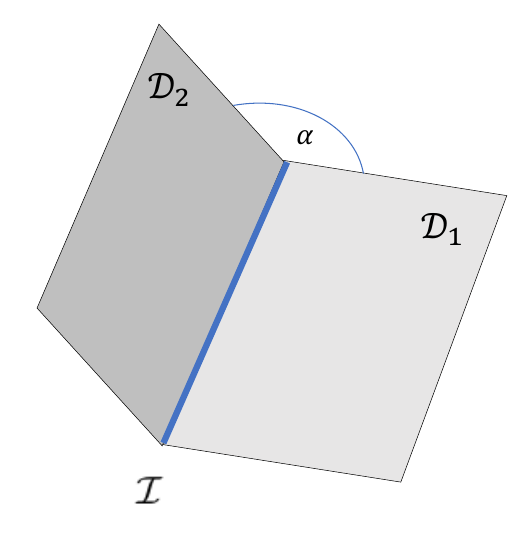}\quad  
    \includegraphics[width=0.27\columnwidth,trim={11cm 5cm 12cm 5cm},clip]{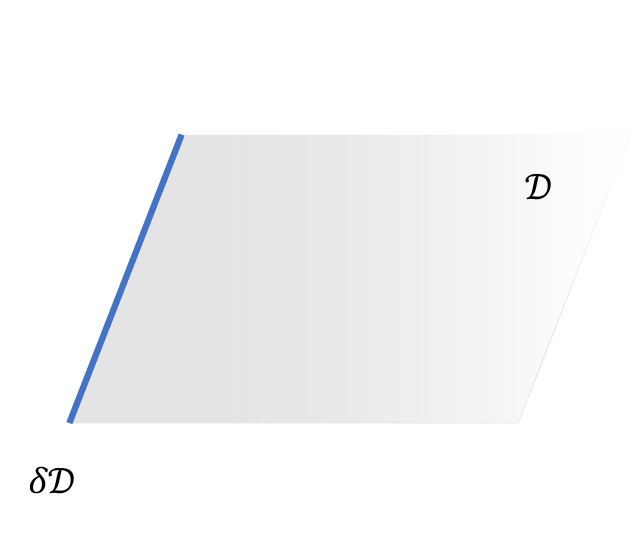}
	\caption{Examples of the defect geometries explored in section \ref{SEC2}. The edge $\delta\mathcal{D}$ or the intersection $\mathcal{I}$ are marked with a thick blue line, and $\alpha$ is the intersection angle. For the wedge (middle picture), the intersection is also the common edge of the two planes.}
	\label{fig:sec2intro}
\end{figure}

In this paper we inquire about what happens when defects meet. We will be able to answer this question and compute new quantities by using the OPE, combined with conformal perturbation theory. Qualitatively, divergences arise when operators from two or more different defects collide at their intersection. These divergences will drive an RG flow localized to the intersection and consequently augment the degrees of freedom in that region. Specifically, it is possible to generate perturbative flows if an operator with a scaling dimension close to the dimension of the intersection is present in the OPE of the two defects.
More so, the same effect happens on a single defect when operators collide on its boundary, which to avoid possible confusion we refer to as its edge. There is therefore an abundance of different geometric constructs that may be considered, since the intersection can serve as the edge of some or all of the defects in the picture. In this manuscript we shall explore among the rest the planar geometries depicted in Fig. \ref{fig:sec2intro}. CFT on the wedge has been explored in the older literature \cite{Cardy1988,Henkel1989,PhysRevD.20.3063,Cardy1983,Barber1984,Guttmann1984,Cardy1984,Bariev1986,Larsson1986,Saxena1987,Wang1990,Igloi1993,Pleimling,PhysRevB.59.65,Pleimling2002}, as well as in contemporary one under broader contexts  \cite{Crampe2013,Antunes:2021qpy,Bissi:2022bgu,Drukker2022,2024arXiv241104043G,Diatlyk:2024zkk,Diatlyk:2024qpr,Kravchuk:2024qoh,Cuomo:2024psk,Geng2021,Miao:2024ddp,Sun:2024qhv,Lai:2024inw}. We also mention works done on conformal interfaces and RG defects \cite{Bachas2002,Fredenhagen2005,PhysRevD.71.066003,Brunner2008,Gaiotto2012,Konechny2013,Konechny2015,Poghosyan2015,Brunner2016,Konechny2017,Konechny2021,Brax:2023goj,10.21468/SciPostPhysCore.7.2.021,Giombi:2024qbm}. 

To exemplify this idea we study perturbative flows in the tricritical model in ${d=3-\epsilon}$ dimensions. For illustration, consider the action (see Fig. \ref{fig:sec2intro}),
\begin{equation}
	S=\int d^{3-\epsilon}x\,\bigl(\frac{1}{2}\left(\partial\phi\right)^{2}+\lambda\phi^{6}\bigr)+g^{\left(1\right)}\int_{\mathcal{D}_{1}}d^{2-\epsilon}x\,\phi^{4}+g^{\left(2\right)}\int_{\mathcal{D}_{2}}d^{2-\epsilon}x\,\phi^{4}+h\int_{\mathcal{I}}d^{1-\epsilon}x\,\phi^{2}
\end{equation}
We study this type of actions for the 3 geometries shown in Fig. \ref{fig:sec2intro} and for two different sets of defect dimensions for both interacting and non-interacting bulk. All of the results are tabulated and discussed in part \ref{tricritical}. Among other things, we shall discuss the interesting dependence of the critical edge coupling $h^*$ on the intersection angle $\alpha$.

\begin{figure}[h!]
	\centering
	\includegraphics[width=0.27\columnwidth,trim={11cm 5cm 12cm 5cm},clip]{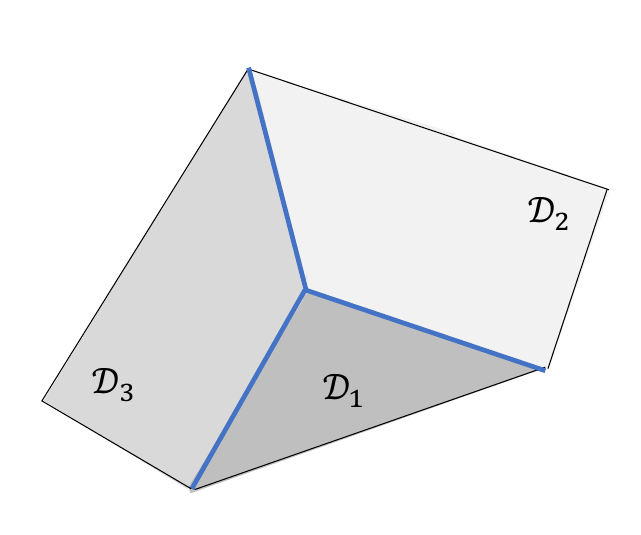}\quad
	\includegraphics[width=0.27\columnwidth,trim={8cm 2.5cm 9cm 3cm},clip]{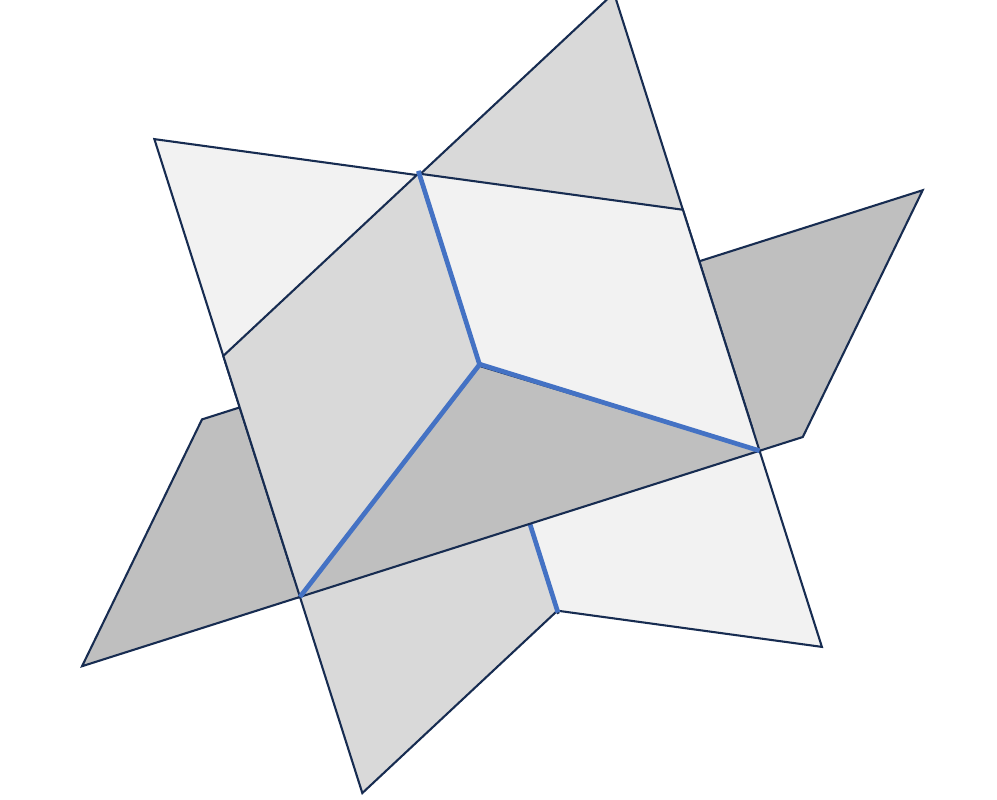}\quad
	\includegraphics[width=0.27\columnwidth,trim={11cm 5cm 12cm 5cm},clip]{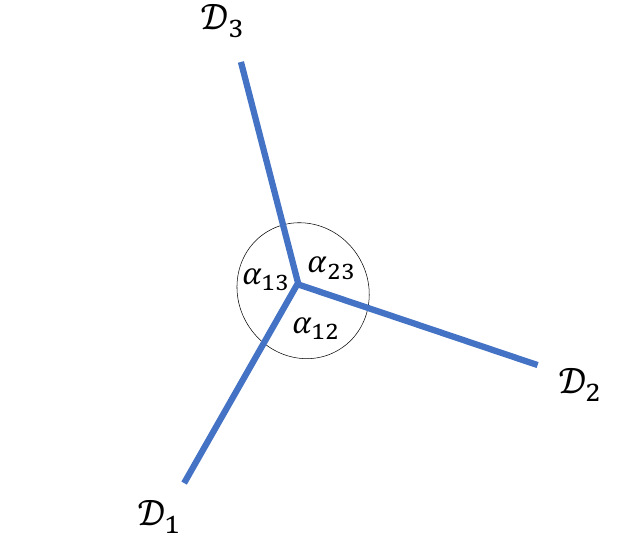}
	\caption{The objects studied in section \ref{SEC3}. \textit{Left:} Trihedral Corner. \textit{Middle:} Extended Trihedral Corner. \textit{Right:} 3-line Corner. The planes meet at a mutual point, giving rise to the corner anomaly depending upon the three relative angles of the edges, $\alpha_{12},\alpha_{23},\alpha_{13}$.}
	\label{fig:sec3intro}
\end{figure}

We may even proceed further and explore more complicated objects, such as three mutually intersecting defects. Fortunately, conformal perturbation analysis is still applicable when the mutual intersection of the three defects is a point. Namely, a corner, as portrayed in Fig. \ref{fig:sec3intro}. At criticality, a central feature of this object is the anomaly in the scaling symmetry around the fixed corner point, which originates from the geometric singularity at the corner. One of the most important results is obtained for the trihedral corner, a corner formed by three intersecting 2-dimensional planes. The defects are defined by the deformations ${\underset{a=1}{\overset{3}{\sum}}g^{\left(a\right)}\int_{\mathcal{D}_{a}}d^{2}x\,\mathcal{O}_{a}}$, and for weak couplings the anomaly arises from a specific term in the cubic order through a logarithmic divergence in the free energy,
\begin{equation} \label{anomintro}
	\log Z=-\Gamma_{\text{corner}}\log\left(\frac{L}{a}\right)\dots=-g^{\left(1\right)}g^{\left(2\right)}g^{\left(3\right)}\int_{\mathcal{D}_{1}}\int_{\mathcal{D}_{2}}\int_{\mathcal{D}_{3}}\left\langle \mathcal{O}_{1}\mathcal{O}_{2}\mathcal{O}_{3}\right\rangle +\dots
\end{equation}
For the extended trihedral corner (see Fig. \ref{fig:sec3intro}), the cubic order gives the leading contribution to the anomaly, which we deem as the corner anomalous dimension, $\Gamma_\text{corner}$. It can be considered therefore as a higher dimensional analog of the cusp anomalous dimension \cite{Polyakov1980,Grozin2016,Cuomo:2024psk}.
\\
For the extended version of the trihedral corner discussed in part \ref{subsec-trihedral} we find,
\begin{equation}
	\Gamma_{\text{corner}}\left(\alpha_{12},\alpha_{23},\alpha_{13}\right)=\frac{\sin\left(\alpha_{12}\right)\sin\left(\alpha_{23}\right)\sin\left(\alpha_{13}\right)}{V^{2}\left(\alpha_{12},\alpha_{23},\alpha_{13}\right)}4\pi^{4}Cg^{\left(1\right)}g^{\left(2\right)}g^{\left(3\right)}
\end{equation}
Where $C$ is the OPE coefficient of the three defect deformations and $V$ is the volume of the unit Parallelepiped entrapped inside the edges of the trihedron,
\begin{equation}
	V^{2}=1+2\cos\left(\alpha_{12}\right)\cos\left(\alpha_{13}\right)\cos\left(\alpha_{23}\right)-\cos^{2}\left(\alpha_{12}\right)-\cos^{2}\left(\alpha_{13}\right)-\cos^{2}\left(\alpha_{23}\right) \label{vol}
\end{equation}

Corners can also be formed by 3 line defects. In such case, the corner contribution \eqref{anomintro} to the anomalous dimension will be sub-leading to the cusp anomalous dimension that arises beforehand at second order,
\begin{equation} 
	\Gamma=\underset{a<b}{\overset{3}{\sum}}\Gamma_{\text{cusp}}\left(\alpha_{ab}\right)+\Gamma_{\text{3-line}}\left(\alpha_{12},\alpha_{23},\alpha_{13}\right)
\end{equation}
We may view each line defect as the world-line of a point-like impurity via conformal mapping to the cylinder $\mathbb{R}\times S^{d-1}$ \cite{Henkel1989,Cuomo:2024psk}. In this interpretation, we can think of $\Gamma_{\text{3-line}}$ as the three-body potential of impurities living on $S^{d-1}$. In part \ref{sec-3l} we find an analytic closed-form expression for this object involving elliptic integrals, and study some of its properties.\\ \\
The rest of the paper is organized as follows: In section \ref{SEC2} we use OPE to derive the running couplings and beta functions for a semi-infinite planar defect \ref{SEC-semi-planes}, two intersecting planes \ref{SEC-intersecting-planes} and a wedge composed of two semi-infinite planes \ref{subsec wedge}. In part \ref{tricritical} we finish with an example of the tricritical model in $d=3-\epsilon$ with $\phi^4$ and $\phi^2$ interactions localized to the planes and the edge respectively. In section \ref{SEC3} we introduce and briefly discuss the corner anomaly. In part \ref{subsec-trihedral} we define the trihedral corner and compute the corner anomalous dimension for its extended version. Subsequently, in part \ref{sec-3l} the 3-line potential is defined and fully computed. Finally, we conclude the paper with a discussion.

\section{Two Defects: RG on the Edge} \label{SEC2}
In this section we use scalar primary operators of the bulk CFT algebra to construct three different geometric setups involving planar defects (see Fig. \ref{fig:sec2intro}). We use the OPE technique \cite{Zamolodchikov:1987ti,Fredenhagen2006,Gaberdiel2009,Fredenhagen2009} to derive the beta functions of the edge and intersection couplings for all cases. Due to the local nature of UV divergences, the results obtained also hold for non-planar defects. Non-flat intersections will have additional curvature terms generated in the RG flow, and their derivation is left outside the scope of this paper.
\subsection{Semi-Infinite Plane} \label{SEC-semi-planes}
Consider a $p-$dimensional semi-infinite planar defect $\mathcal{D}$ embedded in $\mathbb{R}^d$. Coordinates are chosen such that the defect is parameterized by the first p coordinates, with its edge at $x_p=0$,
\begin{equation}
	\mathcal{D}:\left(x_{1},..,x_{p},0,0,..,0\right),\quad x_p\geq 0
\end{equation}
We write the following bare action,
\begin{equation} \label{semi-action}
	S=S_{\text{CFT}}+g_{i,0}\int_{\mathcal{D}}d^{p}x\,\mathcal{O}_{i}\left(\vec{x}_{p-1},x_{p}\right)+h_{\alpha,0}\int_{\mathcal{\partial\mathcal{D}}}d^{p-1}x\,\mathcal{O}_{\alpha}\left(\vec{x}_{p-1},0\right)
\end{equation}
Where for brevity we denoted $\vec{x}_{p-1}=\left(x_{1},..,x_{p-1}\right)$. For future convenience, we use Latin indices to list the set of operators deforming the defect, and Greek indices for the ones on its edge, $\delta\mathcal{D}$. \newline In order to determine the RG flow perturbatively, we expand \eqref{semi-action},
\begin{align} \label{expandsemi}
	e^{-S}=e^{-S_{CFT}}&\left(1-g_{i,0}\int_{\mathcal{D}}d^{p}x\,\mathcal{O}_{i}-h_{\alpha,0}\int_{\mathcal{\partial\mathcal{D}}}d^{p-1}x\,\mathcal{O}_{\alpha}+\frac{1}{2}g_{i,0}g_{j,0}\int_{\mathcal{D}}\int_{\mathcal{D}}\mathcal{O}_{i}\mathcal{O}_{j}\right. \notag \\
	&\;\;\,+\frac{1}{2}h_{\alpha,0}h_{\beta,0}\int_{\mathcal{\partial\mathcal{D}}}\int_{\mathcal{\partial\mathcal{D}}}\mathcal{O}_{\alpha}\mathcal{O}_{\beta}+h_{\alpha,0}g_{i,0}\int_{\mathcal{D}}\int_{\mathcal{\partial\mathcal{D}}}\mathcal{O}_{i}\mathcal{O}_{\alpha}+\dots
\end{align}
A novelty of this construction is a divergent boundary contribution that appears in the $g^2\mathcal{\int_{\mathcal{D}}\int_{\mathcal{D}}O}_{i}\mathcal{O}_{j}$ term which contributes to the renormalization of the edge couplings $h_\alpha$. To see this, we perform an OPE of $\mathcal{O}_{i}\left(x\right)\mathcal{O}_{j}\left(y\right)$ around the point $x$ and regularize the integral over $y$ by removing a small cylindrical "pillbox" of size $\mu^{-1}$ around  point $x$,
\begin{align} \label{semimid}
	&\int_{\mathcal{D}}\int_{\mathcal{D}}\mathcal{O}_{i}\left(x\right)\mathcal{O}_{j}\left(y\right)= \notag \\
	&=\int d^{p-1}x\int^{\left|\vec{y}_{p-1}-\vec{x}_{p-1}\right|\ge\mu^{-1}}d^{p-1}y\int_{x_{p}>0}\,dx_{p}\int_{y_{p}>0}^{\left|y_{p}-x_{p}\right|\ge\mu^{-1}}\,dy_{p}\frac{C_{ij}^{k}\mathcal{O}_{k}\left(\vec{x}_{p-1},x_{p}\right)}{\left|x-y\right|^{\Delta_{ijk}}}+\dots
\end{align}
Where $C_{ij}^{k}$ are the OPE coefficients of the bulk CFT and $\Delta_{ijk}=\Delta_{i}+\Delta_{j}-\Delta_{k}$.\footnote{Indices will be freely raised and lowered with $\delta_{ij}$ without changing the value of the OPE coefficients.} For slightly relevant operators with dimensions $\Delta_{i},\Delta_{j},\Delta_{k}$ all close to $p$, this leads to the usual logarithmic divergence that sets the running of the defect couplings $g_i$. Additionally, when $x_{p}<\mu^{-1}$, the removed pillbox is truncated at the edge at $y_p=0$ and gives an additional boundary contribution. It is obtained from \eqref{semimid} by first integrating over $\vec{y}_{p-1}$ with formula \eqref{form1}, and then taking the lower bound on $y_p=0$ of the remaining integral over $y_p$ that gives,
\begin{align} \label{semihalfway}
	&\int_{\mathcal{D}}\int_{\mathcal{D}}\mathcal{O}_{i}\left(x\right)\mathcal{O}_{j}\left(y\right)= \notag \\
	&=\left(-\frac{\pi^{\frac{p-1}{2}}\Gamma\left(\frac{\Delta_{ijk}-p+1}{2}\right)}{\left(\Delta_{ijk}-p\right)\Gamma\left(\frac{\Delta_{ijk}}{2}\right)}\right)\times\int d^{p-1}x\underset{0}{\overset{\infty}{\int}}dx_{p}\frac{C_{ij}^{k}\mathcal{O}_{k}\left(\vec{x}_{p-1},x_{p}\right)}{\left|x_{p}\right|^{\Delta_{ijk}-p}}\,+\dots
\end{align}
It is evident from \eqref{semihalfway} that divergences can appear as the operator approaches the edge $x_p \rightarrow0$. We perform a partial Taylor expansion of the operator $\mathcal{O}_{k}$ around $x_p =0$ and keep the zero order term,\footnote{An alternative derivation can be done by performing the OPE around the point $x=\left(\vec{x}_{p-1},0\right)$.}
\begin{align} \label{semihalfway2}
	&\int_{\mathcal{D}}\int_{\mathcal{D}}\mathcal{O}_{i}\left(x\right)\mathcal{O}_{j}\left(y\right)= \notag \\
	&=\left(-\frac{\pi^{\frac{p-1}{2}}\Gamma\left(\frac{\Delta_{ij\alpha}-p+1}{2}\right)}{\left(\Delta_{ijk}-p\right)\Gamma\left(\frac{\Delta_{ij\alpha}}{2}\right)}\underset{\mu^{-1}}{\overset{\infty}{\int}}dx_{p}\frac{1}{\left|x_{p}\right|^{\Delta_{ij\alpha}-p}}\right)\times\int d^{p-1}x\,C_{ij}^{\alpha}\mathcal{O}_{\alpha}\left(\vec{x}_{p-1},0\right)+\dots
\end{align}
Notice that now that the operator in \eqref{semihalfway2} is set on the edge we have switched the index $k$ to $\alpha$. We see that divergences will arise when operators with dimension ${\Delta_{\alpha}<p+1-\Delta_{i}-\Delta_{j}}$ are present in the OPE of $\mathcal{O}_{i}\mathcal{O}_{j}$. We focus on perturbative flows, assuming slightly relevant deformations  with dimensions $\Delta_{i}=p-\epsilon_{i}$ and $\Delta_{\alpha}=p-1-\epsilon_{\alpha}$. Other than \eqref{semihalfway2}, the divergences contributing to the renormalization of $h_\alpha$ come from the other terms in \eqref{expandsemi} and are not unusual (see \textit{e.g.} \cite{Shachar2024}).\footnote{There is a factor of half in the  $\int_{\mathcal{D}}\int_{\mathcal{\partial\mathcal{D}}}\mathcal{O}_{i}\mathcal{O}_{\alpha}$ contribution to \eqref{hren-semi} because the planar defect $\mathcal{D}$ is semi-infinite.} In conclusion, we define the following renormalized coupling constant,
\begin{align} \label{hren-semi}
	h_{\alpha}\mu^{\epsilon_{\alpha}}=&h_{\alpha,0}-\frac{\pi^{\frac{p-1}{2}}}{\Gamma\left(\frac{p-1}{2}\right)}\frac{C_{\alpha}^{\beta\gamma}\mu{}^{-\epsilon_{\alpha}}}{\epsilon_{\alpha}}h_{\beta,0}h_{\gamma,0}-\frac{C_{\alpha}^{i\beta}\mu^{-\epsilon_{i}}}{\epsilon_{i}}\frac{\pi^{\frac{p}{2}}}{\Gamma\left(\frac{p}{2}\right)}g_{i,0}h_{\beta,0}+\frac{\pi^{\frac{p-1}{2}}}{2\Gamma\left(\frac{p+1}{2}\right)}\frac{C_{\alpha}^{ij}\mu{}^{-\epsilon_{ij\alpha}}}{\epsilon_{ij\alpha}}g_{i,0}g_{j,0}\notag \\
	&+\mathcal{O}\left(h^{3},gh^{2},g^{2}h,g^{3}\right)
\end{align}
Leading to the following beta function for the edge couplings,
\begin{equation} \label{beta-semi}
	\beta_{\alpha}=-\epsilon_{\alpha}h_{\alpha}+\frac{\pi^{\frac{p-1}{2}}C_{\alpha}^{\beta\gamma}}{\Gamma\left(\frac{p-1}{2}\right)}h_{\beta}h_{\gamma}+\frac{\pi^{\frac{p}{2}}C_{\alpha}^{i\beta}}{\Gamma\left(\frac{p}{2}\right)}g_{i}h_{\beta}-\frac{\pi^{\frac{p-1}{2}}C_{\alpha}^{ij}}{2\Gamma\left(\frac{p+1}{2}\right)}g_{i}g_{j}
\end{equation}
Finally, the beta functions for the couplings $g_i$ are standard and independent of \eqref{beta-semi},
\begin{equation} \label{beta-standard-semi}
	\beta_{i}=-\epsilon_{i}g_{i}+\frac{\pi^{\frac{p}{2}}C_{i}^{jk}}{\Gamma\left(\frac{p}{2}\right)}g_{j}g_{k}
\end{equation}

\subsection{Two Intersecting Planes} \label{SEC-intersecting-planes}
Consider two intersecting planar defects $\mathcal{D}_1,\mathcal{D}_2$ embedded in $\mathbb{R}^d$. We denote their intersection by $\mathcal{I}=\mathcal{D}_{1}\cap\mathcal{D}_{2}$, having dimension $\textit{dim}\left(\mathcal{I}\right)=p-1$. We choose a set of Cartesian coordinates such that $\mathcal{I}$ lies in the first $p-1$ coordinates, and in the $\left(x_{p},x_{p+1}\right)$ space the two planes are tilted by a relative angle $\alpha$ with respect to each other. For greater generality, we also allow one of the planes, arbitrarily chosen to be $\mathcal{D}_2$, to extend in $r$ additional transverse directions. Overall, ${\textit{dim}\left(\mathcal{D}_{1}\right)=p,\; \textit{dim}\left(\mathcal{D}_{2}\right)=p+r}$, and the defects are parameterized as follows,
\begin{align} \label{intersecting-setup}
	&\mathcal{D}_{1}:\left(x_{1},..,x_{p-1},x_{p},0,0,..,0\right) \notag \\
	&\mathcal{D}_{2}:\left(y_{1},..,y_{p-1,}y_{p}\cos\alpha,y_{p}\sin\alpha,y_{p+1}..,y_{p+r},0,0,..,0\right) 
\end{align}
Given such choice, the distance between two points $x\in\mathcal{D}{}_{1},\,y\in\mathcal{D}{}_{2}$ is given by,
\begin{equation} \label{dist}
	\left|x-y\right|^{2}=\vec{y}_{r}^{2}+\left(\vec{x}_{p-1}-\vec{y}_{p-1}\right)^{2}+\left(x_{p}-y_{p}\cos\alpha\right)^{2}+y_{p}^{2}\sin^{2}\alpha
\end{equation}
Where for brevity we denoted $\vec{y}_{p-1}=\left(y_{1},..,y_{p-1}\right)$ and $\vec{y}_{r}=\left(y_{p+1},..,y_{p+r}\right)$. As we anticipate that interactions localized on $\mathcal{I}$ will be generated in the RG flow, we write the following general bare action, 
\begin{equation} \label{inter-action}
	S=S_{\text{CFT}}+g_{i,0}^{\left(1\right)}\int_{\mathcal{D}_{1}}d^{p}x\,\mathcal{O}_{i}\left(x\right)+g_{i,0}^{\left(2\right)}\int_{\mathcal{D}_{2}}d^{p+r}y\,\mathcal{O}_{i}\left(y\right)+h_{\alpha,0}\int_{\mathcal{I}}d^{p-1}x\,\mathcal{O}_{\alpha}\left(\vec{x}_{p-1},0\right)
\end{equation}
Where for future convenience, we use Latin indices to list the operators deforming the two defects and Greek indices for ones on the intersection. To determine the RG flow, we now expand,
\begin{equation} \label{actionExp}
	e^{-S}=e^{-S_{\text{CFT}}}\left(1-\delta S_{1}-\delta S_{2}-\delta S_{\mathcal{I}}\right)
\end{equation}
Where
\begin{align} \label{ds12}
	&\delta S_{1}=g_{i,0}^{\left(1\right)}\int_{\mathcal{D}_{1}}d^{p}x\,\mathcal{O}_{i}-\frac{1}{2}g_{i,0}^{\left(1\right)}g_{j,0}^{\left(1\right)}\int_{\mathcal{D}_{1}}\int_{\mathcal{D}_{1}}\mathcal{O}_{i}\mathcal{O}_{j}+\dots \\
	&\delta S_{2}=g_{i,0}^{\left(2\right)}\int_{\mathcal{D}_{2}}d^{p+r}y\,\mathcal{O}_{i}-\frac{1}{2}g_{i,0}^{\left(2\right)}g_{j,0}^{\left(2\right)}\int_{\mathcal{D}_{2}}\int_{\mathcal{D}_{2}}\mathcal{O}_{i}\mathcal{O}_{j}+\dots \notag
\end{align}
\begin{align} \label{dsI}
\delta S_{\mathcal{I}}=&h_{\alpha,0}\int_{\mathcal{I}}d^{p-1}x\,\mathcal{O}_{\alpha}-\frac{1}{2}h_{\alpha,0}h_{\beta,0}\int_{\mathcal{I}}\int_{\mathcal{I}}\mathcal{O}_{\alpha}\mathcal{O}_{\beta} \\
&-h_{\alpha,0}g_{i,0}^{\left(1\right)}\int_{\mathcal{D}_{1}}\int_{\mathcal{I}}\mathcal{O}_{\alpha}\mathcal{O}_{i}-h_{\alpha,0}g_{i,0}^{\left(2\right)}\int_{\mathcal{D}_{2}}\int_{\mathcal{I}}\mathcal{O}_{\alpha}\mathcal{O}_{i}-g_{i,0}^{\left(1\right)}g_{j,0}^{\left(2\right)}\int_{\mathcal{D}_{1}}\int_{\mathcal{D}_{2}}\mathcal{O}_{i}\mathcal{O}_{j}+\dots \notag
\end{align}
Most of the divergences are standard, with an exception of a novel divergence that arises in the term $\int_{\mathcal{D}_{1}}\int_{\mathcal{D}_{2}}\mathcal{O}_{i}\mathcal{O}_{j}$ that occurs when the two operators collide at the intersection. We explore this in more detail by performing an OPE of $\mathcal{O}_{i}\left(x\right)\mathcal{O}_{j}\left(y\right)$ around the point $x=\rbr{\vec{x}_{p-1},0}$,
\begin{equation} \label{OPE1}
	\int_{\mathcal{D}_{1}}\int_{\mathcal{D}_{2}}\mathcal{O}_{i}\mathcal{O}_{j}=\int d^{p-1}x\int d^{p-1}y\int dx_{p}\int dy_{p}\int d^{r}y\frac{C_{ij}^{\alpha}\mathcal{O}_{\alpha}\left(\vec{x}_{p-1},0\right)}{\left|x-y\right|^{\Delta_{ij\alpha}}}+\dots
\end{equation}
By writing $\left|x-y\right|$ explicitly with \eqref{dist} and using \eqref{form1}, we first integrate over $\vec{y}_{r},\vec{x}_{p-1},x_{p}$, and then introduce a UV cutoff to the resulting integral.\footnote{In principle, we could regulate the integrals over $\vec{y}_{r},\vec{x}_{p-1},x_{p}$ as well, but in the scheme that we use we simply integrate these coordinate away. The universal logarithmic divergences are not sensitive to the choice of regularization scheme.} We get,
\begin{equation} \label{OPE2}
	\int_{\mathcal{D}_{1}}\int_{\mathcal{D}_{2}}\mathcal{O}_{i}\mathcal{O}_{j}=\left(\frac{2\pi^{\frac{p+r}{2}}}{\sin\alpha}\frac{\Gamma\left(\frac{\Delta_{ij\alpha}-p-r}{2}\right)}{\Gamma\left(\frac{\Delta_{ij\alpha}}{2}\right)}\underset{\mu^{-1}}{\overset{\infty}{\int}}dy\frac{1}{y^{\Delta_{ij\alpha}-p-r}}\right)\times\int d^{p-1}x\,C_{ij}^{\alpha}\mathcal{O}_{\alpha}\left(\vec{x}_{p-1},0\right)+\dots
\end{equation}
We focus on perturbative flows, assuming slightly relevant operators deforming $\mathcal{D}_1$ with dimensions  $\Delta_{i}=p-\epsilon_{i}$ and likewise for $\mathcal{D}_2$ with dimensions $\Delta_{j}=p+r-\epsilon_{j}$. Divergent contribution appears in \eqref{OPE2} for operators satisfying $\Delta_{\alpha}<\Delta_{i}+\Delta_{j}-p-r-1$, and as consequence weakly relevant deformations on $\mathcal{I}$ will be turned on for operators with $\Delta_{\alpha}=p-1-\epsilon_{\alpha}$, if they appear in the OPE of $\mathcal{O}_{i}\mathcal{O}_{j}$. Thus we define the renormalized couplings,
\begin{align} \label{inter-renorm}
	h_{\alpha}\mu^{\epsilon_{\alpha}}=h_{\alpha,0}&-\frac{\pi^{\frac{p-1}{2}}}{\Gamma\left(\frac{p-1}{2}\right)}\frac{C_{\alpha}^{\beta\gamma}\mu{}^{-\epsilon_{\alpha}}}{\epsilon_{\alpha}}h_{\beta,0}h_{\gamma,0}-\frac{C_{\alpha}^{\beta i}\mu^{-\epsilon_{i}}}{\epsilon_{i}}h_{\beta,0}\left(\frac{2\pi^{\frac{p}{2}}}{\Gamma\left(\frac{p}{2}\right)}g_{i,0}^{\left(1\right)}+\frac{2\pi^{\frac{p+r}{2}}}{\Gamma\left(\frac{p+r}{2}\right)}g_{i,0}^{\left(2\right)}\right) \notag \\
	&-\frac{1}{\Gamma\left(\frac{p+r+1}{2}\right)}\frac{2\pi^{\frac{p+r+1}{2}}}{\sin\alpha}\frac{C_{\alpha}^{ij}\mu{}^{-\epsilon_{ij\alpha}}}{\epsilon_{ij\alpha}}g_{i,0}^{\left(1\right)}g_{j,0}^{\left(2\right)}+\dots
\end{align}
That lead to the following beta function,
\begin{align} \label{beta-intersecting}
	\beta_{\alpha}=-\epsilon_{\alpha}h_{\alpha}&+\frac{\pi^{\frac{p-1}{2}}C_{\alpha}^{\beta\gamma}}{\Gamma\left(\frac{p-1}{2}\right)}h_{\beta}h_{\gamma}+C_{\alpha}^{\beta i}h_{\beta}\left(\frac{2\pi^{\frac{p}{2}}}{\Gamma\left(\frac{p}{2}\right)}g_{i}^{\left(1\right)}+\frac{2\pi^{\frac{p+r}{2}}}{\Gamma\left(\frac{p+r}{2}\right)}g_{i}^{\left(2\right)}\right)+ \notag \\
	&+\frac{C_{\alpha}^{ij}}{\Gamma\left(\frac{p+r+1}{2}\right)}\frac{2\pi^{\frac{p+r+1}{2}}}{\sin\alpha}g_{i}^{\left(1\right)}g_{j}^{\left(2\right)}.
\end{align}
Finally, independently of \eqref{beta-intersecting} the beta functions for the defect coupling constants $g^{\left(1\right)},g^{\left(2\right)}$ assume the standard form,
\begin{equation} \label{beta-standard-intersecting}
	\beta_{i}^{\left(1\right)}=-\epsilon_{i}g_{i}^{\left(1\right)}+\frac{\pi^{\frac{p}{2}}C_{i}^{jk}}{\Gamma\left(\frac{p}{2}\right)}g_{j}^{\left(1\right)}g_{k}^{\left(1\right)},\qquad \beta_{i}^{\left(2\right)}=-\epsilon_{i}g_{i}^{\left(2\right)}+\frac{\pi^{\frac{p+r}{2}}C_{i}^{jk}}{\Gamma\left(\frac{p+r}{2}\right)}g_{j}^{\left(2\right)}g_{k}^{\left(2\right)}.
\end{equation}
\subsection{Wedge Formed by Two Semi-Infinite planes} \label{subsec wedge}
Consider a wedge that is formed by two semi-infinite $p$-dimensional planar defects that intersect at a mutual edge $\mathcal{I}$. This object is in principle a combination of parts \ref{SEC-semi-planes} and \ref{SEC-intersecting-planes}. We use the former  parameterization \eqref{intersecting-setup} to describe the defects,
\begin{align}
	&\mathcal{D}_{1}:\left(x_{1},..,x_{p-1},x_{p},0,0,..,0\right),\quad &&x_{p}\geq0 \notag \\
	&\mathcal{D}_{2}:\left(y_{1},..,y_{p-1,}y_{p}\cos\alpha,y_{p}\sin\alpha,0,0,..,0\right),\quad &&y_{p}\geq0 
\end{align}
Where now the $\left(p-1\right)$-dimensional mutual edge $\mathcal{I}$ is located at $x_{p}=y_{p}=0$. We also refer to the same action as \eqref{inter-action}, only now with a different defect geometry. As in the previous part, we look at the $\int_{\mathcal{D}_{1}}\int_{\mathcal{D}_{2}}\,\mathcal{O}_{i}\mathcal{O}_{j}$ term in \eqref{dsI} and \newline perform the OPE around the point $x=\left(\vec{x}_{p-1},0\right)$,
\begin{align}
	 &\int_{\mathcal{D}_{1}}\int_{\mathcal{D}_{2}}\,\mathcal{O}_{i}\mathcal{O}_{j}=
	 \notag \\
	 &=\int d^{p-1}x\int d^{p-1}y\underset{0}{\overset{\infty}{\int}}dx_{p}\underset{0}{\overset{\infty}{\int}}dy_{p}\frac{C_{ij}^{\alpha}\mathcal{O}_{\alpha}\left(\vec{x}_{p-1},0\right)}{\left(\left(\vec{x}_{p-1}-\vec{y}_{p-1}\right)^{2}+\left(x_{p}-y_{p}\cos\alpha\right)^{2}+y_{p}^{2}\sin^{2}\alpha\right)^{\frac{\Delta_{ij\alpha}}{2}}}+\dots
\end{align}
After applying \eqref{form1} we have,
\begin{align}
	\int_{\mathcal{D}_{1}}\int_{\mathcal{D}_{2}}\,\mathcal{O}_{i}\mathcal{O}_{j}=\frac{\pi^{\frac{p-1}{2}}}{\sin\alpha}\frac{\Gamma\left(\frac{\Delta_{ij\alpha}-p+1}{2}\right)}{\Gamma\left(\frac{\Delta_{ij\alpha}}{2}\right)}&\int d^{p-1}x\underset{0}{\overset{\infty}{\int}}\frac{dy_{p}C_{ij}^{\alpha}\mathcal{O}_{\alpha}\left(\vec{x}_{p-1},0\right)}{y_{p}^{\Delta_{ijk}-p+1}}\times \\
	&\times\left[\left(x_{p}\right)\,_{2}F_{1}\left(\frac{1}{2},\frac{\Delta_{ijk}-p+1}{2};\frac{3}{2};\frac{-x_{p}^{2}}{y_{p}^{2}}\right)\right]_{x_{p}=-y_{p}\cot\alpha}^{x_{p}\rightarrow\infty} \notag
\end{align}
Where $_{2}F_{1}$ is a Hypergeometric function. For the $x_{p}\rightarrow\infty$ limit, we use the asymptotic expansion,
\begin{align} \label{hypergeom-asympt}
	\left(x_{p}\right)\,_{2}F_{1}\left(\frac{1}{2},\frac{\Delta_{ijk}-p+1}{2};\frac{3}{2};-\frac{x_{p}^{2}}{y_{p}^{2}}\right)&=\frac{\sqrt{\pi}\Gamma\left(\frac{\Delta_{ijk}-p}{2}\right)y_{p}}{2\Gamma\left(\frac{\Delta_{ijk}-p+1}{2}\right)}+\mathcal{O}\left(\frac{1}{x_{p}}\right) \notag \\
	&+\left(\frac{y_{p}^{2}}{x_{p}^{2}}\right){}^{\frac{\Delta_{ijk}-p-1}{2}}\left[\frac{x_{p}}{p-1-\Delta_{ijk}+1}+\mathcal{O}\left(\frac{1}{x_{p}}\right)\right]
\end{align}
and take the first term in \eqref{hypergeom-asympt}.\footnote{The second line in \eqref{hypergeom-asympt} contributes to non-universal power law divergences.} Together with the lower limit of integration over $x_{p}$, we overall have,
\begin{equation}
\int_{\mathcal{D}_{1}}d^{2}x\int_{\mathcal{D}_{2}}d^{2}y\,\mathcal{O}_{i}\left(x\right)\mathcal{O}_{j}\left(y\right)=W\left(\alpha\right)\underset{0}{\overset{\infty}{\int}}\frac{dy_{p}}{y_{p}^{\Delta_{ijk}-p}}\times\int d^{p-1}xC_{ij}^{\alpha}\mathcal{O}_{\alpha}\left(\vec{x}_{p-1},0\right)+\dots
\end{equation}
Where,
\begin{equation} \label{W-def}
	W\left(\alpha\right)=\frac{\pi^{\frac{p-1}{2}}}{\sin\alpha}\frac{\Gamma\left(\frac{\Delta_{ij\alpha}-p-1}{2}\right)}{\Gamma\left(\frac{\Delta_{ij\alpha}}{2}\right)}\left[\frac{\sqrt{\pi}\Gamma\left(\frac{\Delta_{ijk}-p}{2}\right)}{2\Gamma\left(\frac{\Delta_{ijk}-p+1}{2}\right)}+\left(\cot\alpha\right)\,_{2}F_{1}\left(\frac{1}{2},\frac{\Delta_{ij\alpha}-p+1}{2};\frac{3}{2};-\cot^{2}\alpha\right)\right]
\end{equation}
If considering weakly relevant deformations with $\Delta_{i}=p-\epsilon_{i},\;\Delta_{\alpha}=p-1-\epsilon_{\alpha}$, then \eqref{W-def} greatly simplifies with the aid of the identity $\,_{2}F_{1}\left(\frac{1}{2},1;\frac{3}{2};-z^{2}\right)=\frac{\arctan\left(z\right)}{z}$, giving,
\begin{equation} \label{wedge-div}
	\int_{\mathcal{D}_{1}}\int_{\mathcal{D}_{2}}\mathcal{O}_{i}\mathcal{O}_{j}=\left(\frac{\pi-\alpha}{\sin\alpha}\frac{\pi^{\frac{p-1}{2}}}{\Gamma\left(\frac{p+1}{2}\right)}\underset{\mu^{-1}}{\overset{\infty}{\int}}\frac{dy_{p}}{y_{p}^{1-\epsilon_{ij\alpha}}}\right)\times\int d^{p-1}x\,C_{ij}^{\alpha}\mathcal{O}_{\alpha}\left(\vec{x}_{p-1},0\right)+\dots
\end{equation}
As a consistency check, note that the free field cusp anomalous dimension \cite{Cuomo:2024psk} can be retrieved by setting $\mathcal{O}_{\alpha}$ to be the identity operator (hence $p=1$). In such case both \eqref{semihalfway2},\eqref{wedge-div} sum together to yield the correct result. Combining the above with \eqref{hren-semi} and \eqref{inter-renorm}, we arrive at the following beta function,
\begin{align} \label{beta-wedge}
	\beta_{\alpha}=-\epsilon_{\alpha}h_{\alpha}&+\frac{\pi^{\frac{p-1}{2}}C_{\alpha}^{\beta\gamma}}{\Gamma\left(\frac{p-1}{2}\right)}h_{\beta}h_{\gamma}+\frac{\pi^{\frac{p}{2}}C_{\alpha}^{i\beta}}{\Gamma\left(\frac{p}{2}\right)}\left(g_{i}^{\left(1\right)}+g_{i}^{\left(2\right)}\right)h_{\beta} \notag \\
	&+\frac{\pi^{\frac{p-1}{2}}C_{\alpha}^{ij}}{\Gamma\left(\frac{p+1}{2}\right)}\left(\frac{\pi-\alpha}{\sin\alpha}g_{i}^{\left(1\right)}g_{j}^{\left(2\right)}-\frac{g_{i}^{\left(1\right)}g_{j}^{\left(1\right)}+g_{i}^{\left(2\right)}g_{j}^{\left(2\right)}}{2}\right).
\end{align}
Note that when $g^{\left(1\right)},g^{\left(2\right)}$ are identical and $\alpha =\pi$, the last term in \eqref{beta-wedge} vanish. Indeed, in this case there is just an infinite plane and the edge disappears.\footnote{More exactly, if $h_\alpha =0$ it will remain so. It is possible to start with a non vanishing $h_\alpha$ that will flow to a non-trivial fixed-point. That would describe a composite defect \cite{Shimamori2024,Ge:2024hei}.}
\subsection{Example: $\left(\phi^{4}\right)_{\mathcal{D}_1}\cap\left(\phi^{4}\right)_{\mathcal{D}_2}:\left(\phi^{2}\right)_{\mathcal{I}}$ in Tricritical Bulk} \label{tricritical}
This is a nice tractable example of a perturbative RG flow that can be studied with the tools that were presented. We consider a bulk theory in $d=3-\epsilon$ dimensions with $N$ scalar fields that is described by the tricritical model,
\begin{equation} \label{sbulk}
	S_{\text{bulk}}=\int d^{3-\epsilon}x\left(\frac{1}{2}\left(\partial\phi\right)^{2}+\lambda\left(\phi^{2}\right)^{3}\right)
\end{equation}
The operators $\phi^{2}$ and $\phi^{4}$ have dimensions close to and 1 and 2 respectively, and are hence available for constructing the different geometric objects studied above. The bulk coupling has the following beta function to leading order in epsilon \cite{PhysRevLett.48.574,PhysRevB.37.5257},
\begin{equation}
	\beta_{\lambda}=-2\epsilon\lambda+\frac{3\left(3N+22\right)}{\pi^{2}}\lambda^{2}
\end{equation}
At the IR fixed point $\lambda^{*}=\frac{2\pi^{2}\epsilon}{3\left(3N+22\right)}$, the operators $\phi^{2}$ and $\phi^{4}$ assume the following anomalous dimensions,
\begin{gather}
	\gamma_{\phi^{4}}=\frac{4\left(N+4\right)}{3N+22}\epsilon \label{anom-dim-phi4} \\
	\gamma_{\phi^{2}}=\frac{5\left(N+2\right)\left(N+4\right)}{4\left(3N+22\right)^{2}}\epsilon^{2} \label{anom-dim-phi2} 
\end{gather}
To utilize conformal perturbation theory in what follows, we use the following data,
\begin{gather} \label{free-CFT-data}
	C_{\phi^{2}}^{\phi^{2}\phi^{2}}=4C_{\phi},\quad C_{\phi^{2}}^{\phi^{4}\phi^{2}}=4\left(N+2\right)C_{\phi}^{2},\quad C_{\phi^{2}}^{\phi^{4}\phi^{4}}=32\left(N+2\right)C_{\phi}^{3}  \\
	C_{\phi^{4}}^{\phi^{4}\phi^{4}}=8\left(N+8\right)C_{\phi}^{2},\quad C_{\phi}=\frac{\Gamma\left(\frac{d-2}{2}\right)}{4\pi^{\frac{d}{2}}}. \notag
\end{gather}
\subsubsection*{Semi-Infinite Plane}
This case is described by the action,
\begin{equation}
	S=S_{\text{bulk}}+g_{0}\int_{\mathcal{D}}d^{p}x\,\phi^{4}+h_{0}\int_{\delta\mathcal{D}}d^{p-1}x\,\phi^{2}
\end{equation}
Where the bulk is fixed in dimension $d=3-\epsilon$ and controlled by action \eqref{sbulk}. We can consider (at least) two different models, which are both interesting on their own. One is for $p=2$ and the second is for $p=2-\epsilon$. We shall inspect both of them. The beta functions are obtained from \eqref{beta-semi},\eqref{beta-standard-semi} and \eqref{free-CFT-data},
\begin{gather}
	\beta_{g}=-\left(p-\Delta_{\phi^{4}}\right)g+\left(N+8\right)\frac{g^{2}}{2\pi} \label{betag-semi-ex} \\
	\beta_{h}=-\left(p-1-\Delta_{\phi^{2}}\right)h+\frac{h^{2}}{\pi}+\frac{N+2}{4}\frac{gh}{\pi}-\frac{N+2}{2}\frac{g^{2}}{\pi^{3}} \label{betah-semi-ex}
\end{gather}
Where $\Delta_{\phi^{4}}=2-2\epsilon,\;\Delta_{\phi^{2}}=1-\epsilon\;$ for non-interacting bulk, and when the bulk is critical at $\lambda=\lambda^*$ these are added with the anomalous dimensions \eqref{anom-dim-phi4},\eqref{anom-dim-phi2}.\footnote{The anomalous dimension \eqref{anom-dim-phi2} is order $\mathcal{O}\left(\epsilon^2\right)$, so does not contribute to the leading order in $\epsilon$.} We now explore the various possibilities,
\begin{itemize}
	\item $p=2,\;\lambda=0$: 
	\begin{gather}
		g^{*}=\frac{4\pi\epsilon}{N+8}
		,\qquad
		h_{\pm}^{*}=\frac{3\pi\epsilon}{N+8}\pm\frac{\pi}{2}\gamma_{e,1} \\
		\gamma_{e,1}=\frac{6\epsilon}{N+8}\sqrt{1+8\frac{N+2}{\left(3\pi\right)^{2}}}
	\end{gather}
	
	\item $p=2-\epsilon,\;\lambda=0$: 
	\begin{gather}
		g^{*}=\frac{2\pi\epsilon}{N+8}
		,\qquad
		h_{\pm}^{*}=-\frac{N+2}{N+8}\frac{\pi\epsilon}{4}\pm\frac{\pi}{2}\gamma_{e,2} \\
		\gamma_{e,2}=\frac{N+2}{N+8}\frac{\epsilon}{2}\sqrt{1+\frac{32}{\pi^{2}\left(N+2\right)}}
	\end{gather}
	
	\item $p=2,\;\lambda=\lambda^*$: 
	\begin{gather}
		g^{*}=\frac{4\left(N+14\right)}{\left(3N+22\right)\left(N+8\right)}\pi\epsilon
		\\
		h_{\pm}^{*}=\frac{\left(N^{2}+15N+74\right)\pi\epsilon}{\left(3N+22\right)\left(N+8\right)}\pm\frac{\pi}{2}\gamma_{e,3}
		\\
		\gamma_{e,3}=\frac{2\left(N^{2}+15N+74\right)\epsilon}{\left(3N+22\right)\left(N+8\right)}\sqrt{1+\frac{8\left(N+2\right)\left(N+14\right)^{2}}{\pi^{2}\left(N^{2}+15N+74\right)^{2}}}
	\end{gather}
	
	\item $p=2-\epsilon,\;\lambda=\lambda^*$: 
	\begin{gather}
		g^{*}=-\frac{N-6}{\left(3N+22\right)\left(N+8\right)}\pi\epsilon
		\\
		h_{\pm}^{*}=\frac{3\left(N-6\right)\left(N+2\right)\pi\epsilon}{24\left(3N+22\right)\left(N+8\right)}\pm\frac{\pi}{2}\gamma_{e,4}
		\\
		\gamma_{e,4}=\frac{6\left(N-6\right)\left(N+2\right)\epsilon}{\left(3N+22\right)\left(N+8\right)}\sqrt{1+\frac{32}{\pi^{2}\left(N+2\right)}}
	\end{gather}
\end{itemize}
In all cases the discriminant is positive for all $N$. The flow takes its course from the UV fixed-point $h_{-}^{*}$ to the IR fixed-point $h_{+}^{*}$. By shifting the coupling constant to $\tilde{h}=h-h_{-}^{*}$, the beta function \eqref{betah-semi-ex} takes the form,
\begin{equation} \label{beta-shifted}
	\beta_{\tilde{h}}=-\gamma_{e}\tilde{h}+\frac{\tilde{h}^{2}}{\pi}
\end{equation}
The interpretation of \eqref{beta-shifted} is simple - the flow begins at the UV from a non-trivial weakly coupled "Edge-CFT", deformed by a slightly relevant edge deformation with dimension $\Delta=p+\gamma_{e}$ and terminating at the IR fixed-point with $\tilde{h}^{*}=h_{+}^{*}-h_{-}^{*}$.
\subsubsection*{Two Intersecting Planes}
We use the following action,
\begin{equation}
	S=S_{\text{bulk}}+g_{0}^{\left(1\right)}\int_{\mathcal{D}_{1}}d^{p}x\,\phi^{4}+g_{0}^{\left(2\right)}\int_{\mathcal{D}_{2}}d^{p}x\,\phi^{4}+h_{0}\int_{\mathcal{I}}d^{p-1}x\,\phi^{2}
\end{equation}
The beta functions are obtained from \eqref{beta-intersecting},\eqref{beta-standard-intersecting} and \eqref{free-CFT-data}. In this case $g^{\left(1\right)}=g^{\left(2\right)}$ and we simply call both of these couplings $g$. Therefore,
\begin{gather}
	\beta_{g}=-\left(p-\Delta_{\phi^{4}}\right)g+\left(N+8\right)\frac{g^{2}}{2\pi} \label{betag-semi-int} \\
	\beta_{h}=-\left(p-1-\Delta_{\phi^{2}}\right)h+\frac{h^{2}}{\pi}+\left(N+2\right)\frac{gh}{\pi}+\left(N+2\right)\frac{g^{2}}{\pi^{2}\sin\alpha} \label{betah-semi-int}
\end{gather} \newpage
We list the four different possibilities,
\begin{itemize}
	\item $p=2,\;\lambda=0,\;g^{*}=\frac{4\pi\epsilon}{N+8}$: 
	\begin{gather}
		h_{\pm}^{*}=-\frac{N-4}{N+8}\frac{\pi\epsilon}{2}\pm\frac{\pi}{2}\gamma_{i,1}\left(\alpha\right) \\
		\gamma_{i,1}\left(\alpha\right)=\frac{N-4}{N+8}\epsilon\sqrt{1-\frac{32\left(N+2\right)}{\pi\left(N-4\right)^{2}\sin\alpha}}
	\end{gather}
	
	\item $p=2-\epsilon,\;\lambda=0,\;g^{*}=\frac{2\pi\epsilon}{N+8}$: 
	\begin{gather}
		h_{\pm}^{*}=-\frac{N+2}{N+8}\pi\epsilon\pm\frac{\pi}{2}\gamma_{i,2}\left(\alpha\right) \\
		\gamma_{i,2}\left(\alpha\right)=2\frac{N+2}{N+8}\epsilon\sqrt{1-\frac{8}{\pi\left(N+2\right)\sin\alpha}}
	\end{gather}
	
	\item $p=2,\;\lambda=\lambda^*,\;g^{*}=\frac{4\left(N+14\right)}{\left(3N+22\right)\left(N+8\right)}\pi\epsilon$: 
	\begin{gather}
		h_{\pm}^{*}=-\frac{\left(N^{2}+18N-64\right)\pi\epsilon}{2\left(N+8\right)\left(3N+22\right)}\pm\frac{\pi}{2}\gamma_{i,3}\left(\alpha\right)
		\\
		\gamma_{i,3}\left(\alpha\right)=\frac{\left(N^{2}+18N-64\right)\epsilon}{4\left(N+8\right)\left(3N+22\right)}\sqrt{1-\frac{128\left(N+2\right)\left(N+14\right)^{2}}{\pi\left(N^{2}+18N-64\right)^{2}\sin\alpha}}
	\end{gather}
	
	\item $p=2-\epsilon,\;\lambda=\lambda^*\;,g^{*}=-\frac{N-6}{\left(3N+22\right)\left(N+8\right)}\pi\epsilon$: 
	\begin{gather}
		h_{\pm}^{*}=\frac{\pi\left(N-6\right)\left(N+2\right)\epsilon}{2\left(N+8\right)\left(3N+22\right)}\pm\frac{\pi}{2}\gamma_{i,4}\left(\alpha\right)
		\\
		\gamma_{i,4}\left(\alpha\right)=\frac{\left(N-6\right)\left(N+2\right)\epsilon}{4\left(N+8\right)\left(3N+22\right)}\sqrt{1-\frac{8}{\pi\left(N+2\right)\sin\alpha}}
	\end{gather}
\end{itemize}
As the intersection angle is gradually decreased, a putative critical angle appears when $\gamma_i\left(\alpha_c\right)=0$. We note that this behavior was also observed in \cite{Cardy1983,Larsson1986} (see also comment in \cite{Wang1990}). At this level of computation, we can not determine if this indicates an actual physical phenomenon or whether this is an artifact of perturbation theory. The latter is definitely plausible, as for when $\gamma_i=0$ one must resort to the next order in perturbation theory.
\subsubsection*{Wedge Formed by Two Semi-Infinite planes}
This is a modification of the above, where the beta functions are obtained from  \eqref{beta-standard-intersecting},\eqref{beta-wedge} and \eqref{free-CFT-data}. In this case $g^{\left(1\right)}=g^{\left(2\right)}$ and we simply call both of these couplings $g$,
\begin{gather}
	\beta_{g}=-\left(p-\Delta_{\phi^{4}}\right)g+\left(N+8\right)\frac{g^{2}}{2\pi} \label{betag-wedge} \\
	\beta_{h}=-\left(p-1-\Delta_{\phi^{2}}\right)h+\frac{h^{2}}{\pi}+\left(N+2\right)\frac{gh}{2\pi}+\left(N+2\right)W\left(\alpha\right)\frac{g^{2}}{\pi^{3}} \label{betah-wedge}
\end{gather}
Where,
\begin{equation}
	W\left(\alpha\right)=\frac{\pi-\alpha}{\sin\left(\alpha\right)}-1
\end{equation}
The four cases are then tabulated as,
\begin{itemize}
	\item $p=2,\;\lambda=0,\;g^{*}=\frac{4\pi\epsilon}{N+8}$: 
	\begin{gather}
		h_{\pm}^{*}=-\frac{N-4}{N+8}\frac{\pi\epsilon}{2}\pm\frac{\pi}{2}\gamma_{w,1}\left(\alpha\right) \\
		\gamma_{w,1}\left(\alpha\right)=\frac{N-4}{N+8}\epsilon\sqrt{1-\frac{64\left(N+2\right)}{\pi^{2}\left(N-4\right)^{2}}W\left(\alpha\right)}
	\end{gather}
	
	\item $p=2-\epsilon,\;\lambda=0,\;g^{*}=\frac{2\pi\epsilon}{\left(N+8\right)}$: 
	\begin{gather}
		h_{\pm}^{*}=-\frac{N+2}{N+8}\frac{\pi\epsilon}{2}\pm\frac{\pi}{2}\gamma_{w,2}\left(\alpha\right) \\
		\gamma_{w,2}\left(\alpha\right)=\frac{N+2}{N+8}\epsilon\sqrt{1-\frac{16W\left(\alpha\right)}{\pi^{2}\left(N+2\right)}}
	\end{gather}
	
	\item $p=2,\;\lambda=\lambda^*,\;g^{*}=\frac{4\left(N+14\right)}{\left(3N+22\right)\left(N+8\right)}\pi\epsilon$: 
	\begin{gather}
		h_{\pm}^{*}=\frac{N^{2}+14N+120}{\left(N+8\right)\left(3N+22\right)}\frac{\pi\epsilon}{2}\pm\frac{\pi}{2}\gamma_{w,3}\left(\alpha\right)
		\\
		\gamma_{w,3}\left(\alpha\right)=\frac{N^{2}+14N+120}{\left(N+8\right)\left(3N+22\right)}\epsilon\sqrt{1-\frac{64\left(N+2\right)\left(N+14\right)^{2}W(\alpha)}{\pi^{2}\left(N^{2}+14N+120\right)^{2}}}
	\end{gather}
	
	\item $p=2-\epsilon,\;\lambda=\lambda^*,\;g^{*}=-\frac{N-6}{\left(3N+22\right)\left(N+8\right)}\pi\epsilon$: 
	\begin{gather}
		h_{\pm}^{*}=\frac{\left(N-6\right)\left(N+2\right)}{\left(N+8\right)\left(3N+22\right)}\frac{\pi\epsilon}{4}\pm\frac{\pi}{2}\gamma_{w,4}\left(\alpha\right)
		\\
		\gamma_{w,4}\left(\alpha\right)=\frac{\left(N-6\right)\left(N+2\right)}{\left(N+8\right)\left(3N+22\right)}\frac{\epsilon}{2}\sqrt{1-\frac{16W(\alpha)}{\pi^{2}\left(N+2\right)}}
	\end{gather}
\end{itemize}
Just like for the two intersecting planes discussed previously, the issue of the putative critical angle that appears as $\alpha$ is decreased remains here as well. The $\alpha=\pi$ case gives the values for the composite line defect.

\section{Three Defects: Trihedral \& 3-Line Corners} \label{SEC3}
We may also explore the possibility of the emergence of degrees of freedom taking place at the mutual intersection of more than two defects. In contrast to two defects nevertheless, novel terms in the beta function such as the one appearing in \eqref{beta-intersecting} begin from cubic order (for 3 defects for instance, there would be a term in the beta function $\propto g^{\left(1\right)}g^{\left(2\right)}g^{\left(3\right)}$) and hence fall beyond the scope of conformal perturbation theory. A notable exception however in when conformal perturbation can be applied beyond two defects is when the mutual intersection is a point.

The minimal number of $p$-dimensional planes that are required to mutually intersect at a point is $p+1$. Indeed, the simplest object of this kind is a cusp formed by two lines ($p=1$), presenting itself in the free energy as a logarithmic term known as the cusp anomalous dimension \cite{Polyakov1980,Grozin2016,Cuomo:2024psk}. Furthermore, we may consider three 2-dimensional planes intersecting at a point ($p=2$), an object that we refer to as a Trihedral Corner. It will give rise to an anomalous dimension through a novel logarithmic term appearing in the free energy that is remarkably still accessible by conformal perturbation theory at the level of the three-point function.

\subsection{Trihedral Corners} \label{subsec-trihedral}
\begin{figure}[h!]
	\centering
	\includegraphics[width=0.35\columnwidth,,trim={11cm 5cm 11cm 4cm},clip]{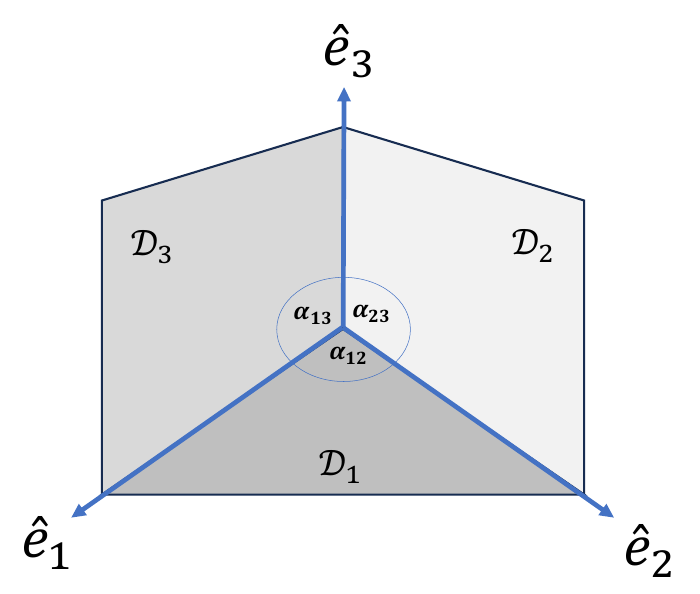}\quad
	\includegraphics[width=0.35\columnwidth,trim={8.5cm 4cm 8.5cm 3cm},clip]{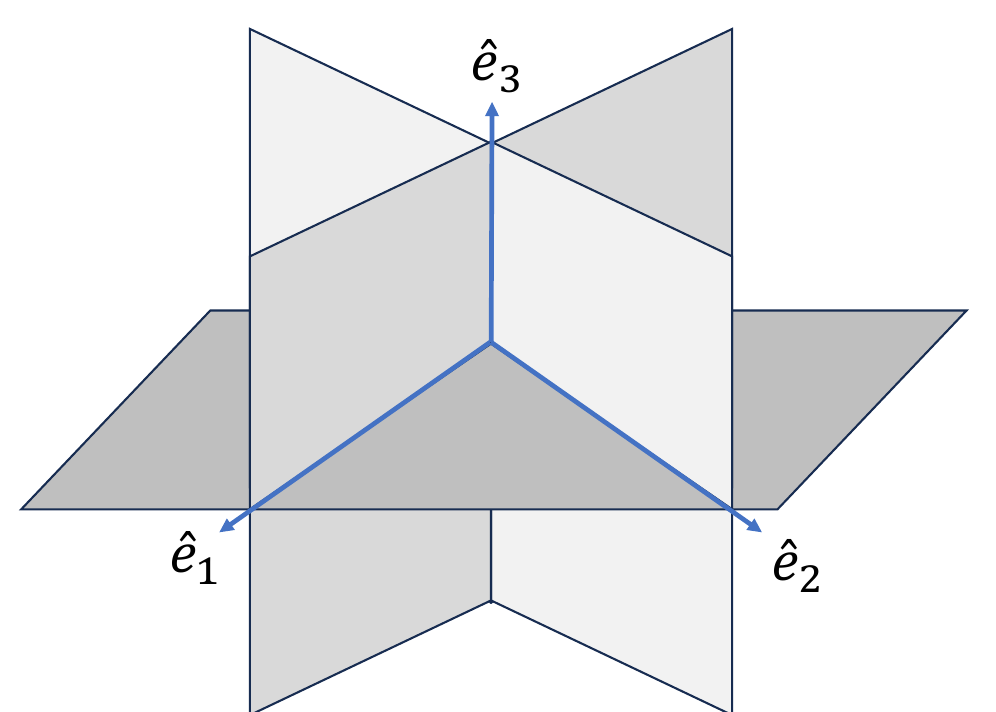}
	\caption{\textit{Left:} A trihedral corner subtended by the three unit vectors $\hat{e}_{1},\hat{e}_{2},\hat{e}_{3}$. The relative angles $\hat{e}_{a}\cdot\hat{e}_{b}=\cos\left(\alpha_{ab}\right)$ are also drawn. \textit{Right:} An extended trihedral corner.}
	\label{fig:trihedral}
\end{figure}

To study the trihedral corner anomaly, consider three 2-dimensional planar defects \linebreak in $\mathbb{R}^{d\geq3}$: $\mathcal{D}_1,\mathcal{D}_2,\mathcal{D}_3$, as they are described by the action, 
\begin{equation} \label{corner action}
	S=S_{\text{CFT}}+\overset{3}{\underset{a=1}{\sum}}g_{i}^{\left(a\right)}\int_{\mathcal{D}_{a}}d^{2}x\,\mathcal{O}_{i}
\end{equation}
As a lesson from the last section, we should also account for the fact that three line defects will possibly be generated at the intersection of each pair of planes. It is important to emphasize that this effect is not directly related to the trihedral corner anomaly, but it is interesting on its own regardless, and its ramifications will be separately discussed in part \ref{sec-3l}.
For the computation of the anomaly coefficient we assume that $\Delta_{i}=2$, although the result is also applicable for slightly relevant deformations. The corner anomalous dimension $\Gamma_{\text{trihedral}}$ will arise from the following term in the free energy,\footnote{$L$ and $a$ are IR and UV regulators respectively. $L$ can be thought of as the size of the defects.}
\begin{align} \label{CAD}
	\log Z&=-\Gamma_{\text{trihedral}}\log\left(\frac{L}{a}\right)+\dots \notag \\
	&=-g_{i}^{\left(1\right)}g_{j}^{\left(2\right)}g_{k}^{\left(3\right)}\int_{\mathcal{D}_{1}}d^{2}x\int_{\mathcal{D}_{2}}d^{2}y\int_{\mathcal{D}_{3}}d^{2}z\left\langle \mathcal{O}_{i}\left(x\right)\mathcal{O}_{j}\left(y\right)\mathcal{O}_{k}\left(z\right)\right\rangle +\dots
\end{align}
We define the three unit vectors $\hat{e}_{1},\hat{e}_{2}$ and $\hat{e}_{3}$ such that they lay parallel to the edges of intersection $\mathcal{I}_{13},\mathcal{I}_{12},\mathcal{I}_{23}$ respectively (see Fig. \ref{fig:trihedral}). The vectors are used to parameterize the three planes by,
\begin{equation} \label{3planes-param}
	\mathcal{D}_{1}\left(x_{1},x_{2}\right)=x_{1}\hat{e}_{1}+x_{2}\hat{e}_{2},\;\mathcal{D}_{2}\left(y_{1},y_{2}\right)=y_{1}\hat{e}_{2}+y_{2}\hat{e}_{3},\;\mathcal{D}_{3}\left(z_{1},z_{2}\right)=z_{1}\hat{e}_{3}+z_{2}\hat{e}_{1}.
\end{equation}
 \newline
There are (at least) two geometric objects that we may consider. The first object is literally a corner. This is the 3 defect analog of the wedge \ref{subsec wedge} and it is parameterized by requiring that all parameters $x_1,x_2,\dots,z_2$ in \eqref{3planes-param} be positive. More accurately, it is the boundary of the polyhedral convex cone generated by $\hat{e}_{1},\hat{e}_{2},\hat{e}_{3}$.\newline The second geometric object is an extended corner. This is the analog of \ref{SEC-intersecting-planes} and can be considered as the gluing of eight corners. To put it differently, all three planes extend indefinitely, with all the parameters in \eqref{3planes-param} unrestricted other than being real.
We shall perform the computation for the extended trihedral corner, where the second case is left for a future study.\footnote{The structure of the non-extend corner is more involved. This is because every single face of the trihedron has a corner independently, and therefore preceding (\ref{CAD}) there will be logarithmic contributions already appearing at the level of the two-point function.} For that end we write the integrated three-point \newline function \eqref{CAD} explicitly by using parameterization \eqref{3planes-param},
\begin{equation} \label{corner int}
	\int_{\mathcal{D}_{1}}\int_{\mathcal{D}_{2}}\int_{\mathcal{D}_{3}}\left\langle \mathcal{O}_{i}\mathcal{O}_{j}\mathcal{O}_{k}\right\rangle =C_{ijk}\int d^{2}x\int d^{2}y\int d^{2}z\frac{\sin\left(\alpha_{12}\right)\sin\left(\alpha_{23}\right)\sin\left(\alpha_{13}\right)}{\left|x-y\right|^{2}\left|x-z\right|^{2}\left|z-y\right|^{2}}
\end{equation}
\begin{align} 
	\left|x-y\right|^{2}=&x_{1}^{2}+y_{2}^{2}+\left(x_{2}-y_{1}\right){}^{2}+2y_{2}\cos\left(\alpha_{23}\right)\left(y_{1}-x_{2}\right) \notag \\
	&+2x_{1}\left(\cos\left(\alpha_{12}\right)\left(x_{2}-y_{1}\right)-y_{2}\cos\left(\alpha_{13}\right)\right). \label{disttrihedral}
\end{align}
Where the other distances in \eqref{corner int} are obtained from permutations of \eqref{disttrihedral}. Next we preform Feynman parameterization \eqref{Feynman} and define the vector $X=\left(x_{1},x_{2},y_{1},y_{2},z_{1},z_{2}\right)$. As so, the integral inside \eqref{corner int} can be expressed as,
\begin{equation}
	\iiint\frac{d^{2}x\,d^{2}y\,d^{2}z}{\left|x-y\right|^{2}\left|x-z\right|^{2}\left|z-y\right|^{2}}=2\underset{0}{\overset{1}{\int}}du\underset{0}{\overset{1}{\int}}dv\left(1-v\right)\int\frac{d^{6}X}{\left(X^{T}MX\right)^{6}}
\end{equation}
Where $M$ is a $\left(6\times6\right)$ real symmetric matrix with coefficients depending on $u,v$ and $\alpha_{12},\alpha_{23},\alpha_{13}$. We diagonalize by change of variables,
\begin{equation}
	\iiint\frac{d^{2}x\,d^{2}y\,d^{2}z}{\left|x-y\right|^{2}\left|x-z\right|^{2}\left|z-y\right|^{2}}=2\pi^{3}\underset{0}{\overset{1}{\int}}du\underset{0}{\overset{1}{\int}}dv\frac{1-v}{\left(detM\left(u,v\right)\right)^{1/2}}\int\frac{dX}{X}
\end{equation}
Remarkably, the determinant factorizes to,
\begin{gather}
	\det M=\left(1-v\right)^{3}\left(u\left(1-u\right)\left(1-v\right)+v\right)^{3}\times V^{4}\left(\alpha_{12},\alpha_{23},\alpha_{13}\right) \notag \\
	V^{2}\left(\alpha_{12},\alpha_{23},\alpha_{13}\right)=1+2\cos\left(\alpha_{12}\right)\cos\left(\alpha_{13}\right)\cos\left(\alpha_{23}\right)-\cos^{2}\left(\alpha_{12}\right)-\cos^{2}\left(\alpha_{13}\right)-\cos^{2}\left(\alpha_{23}\right) \label{vol}
\end{gather}
Where  $V\left(\alpha_{12},\alpha_{23},\alpha_{13}\right)=\left|\hat{e}_{1}\wedge\hat{e}_{2}\wedge\hat{e}_{3}\right|$ is the volume of the unit Parallelepiped subtended by $\hat{e}_{1},\hat{e}_{2},\hat{e}_{3}$. The integrals over $u,v$ can then be carried out straightforwardly and one finds,
\begin{equation}
	\iiint\frac{d^{2}x\,d^{2}y\,d^{2}z}{\left|x-y\right|^{2}\left|x-z\right|^{2}\left|z-y\right|^{2}}=\frac{4\pi^{4}}{V^{2}\left(\alpha_{12},\alpha_{23},\alpha_{13}\right)}\int\frac{dX}{X}
\end{equation}
In conclusion,
\begin{equation} \label{CAD-result}
	\Gamma_{\text{ext.-trihedral}}\left(\alpha_{12},\alpha_{23},\alpha_{13}\right)=\frac{\sin\left(\alpha_{12}\right)\sin\left(\alpha_{23}\right)\sin\left(\alpha_{13}\right)}{V^{2}\left(\alpha_{12},\alpha_{23},\alpha_{13}\right)}4\pi^{4}C^{ijk}g_{i}^{\left(1\right)}g_{j}^{\left(2\right)}g_{k}^{\left(3\right)}
\end{equation} \\
As expected, \eqref{CAD-result} is symmetric with respect to $\alpha_{12},\alpha_{23},\alpha_{13}$ and diverges when $\hat{e}_{1},\hat{e}_{2},\hat{e}_{3}$ are co-planar, meaning an overlap between the planes. Furthermore, a special case that is worth mentioning is when two angles are equal to $\frac{\pi}{2}$ and the remaining one to $\alpha$,
\begin{equation}
	\Gamma_{\text{ext.-trihedral}}\left(\alpha\right)=\frac{4\pi^{4}}{\sin\alpha}C^{ijk}g_{i}^{\left(1\right)}g_{j}^{\left(2\right)}g_{k}^{\left(3\right)}
\end{equation}
Note that this functional dependence on $\alpha$ is identical to the one appearing in the anomalous dimension of two intersecting lines \eqref{CAD 2 int}.

\subsection{The 3-Line Potential}\label{sec-3l}
\begin{figure}[h!]
	\centering
	 \includegraphics[width=0.3\columnwidth,,trim={11cm 5cm 11cm 5cm},clip]{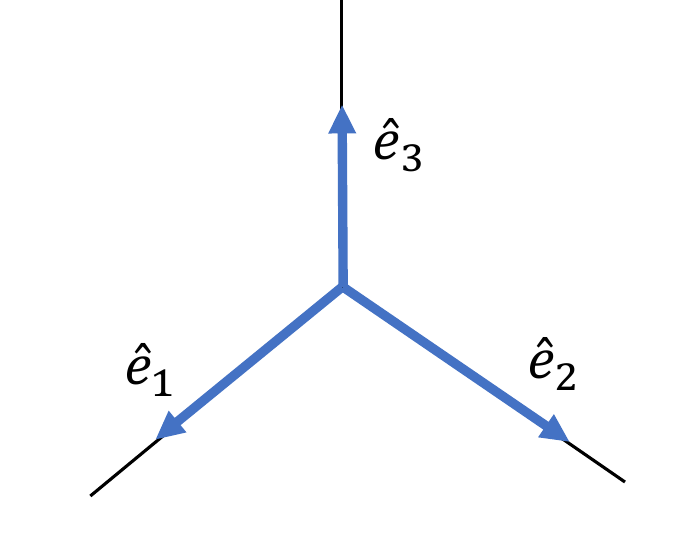}\quad
	 \includegraphics[width=0.3\columnwidth,trim={11cm 5cm 11cm 5cm},clip]{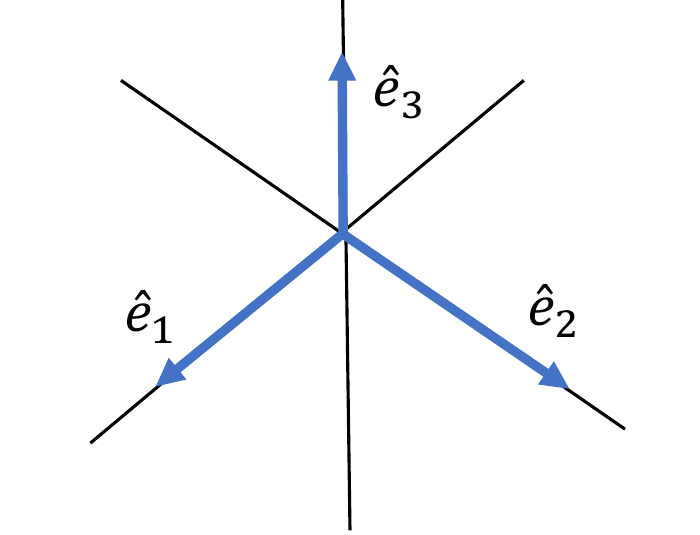}
	\caption{\textit{Left:} A 3-line corner defined with $\hat{e}_{1},\hat{e}_{2},\hat{e}_{3}$ and parameterized by $x,y,z \geq0$. \textit{Right:} An extended 3-line corner, parameterized by $x,y,z \in\mathbb{R}$. Relative angles are defined by $\hat{e}_{a}\cdot\hat{e}_{b}=\cos\left(\alpha_{ab}\right)$.}
	\label{fig:3lcorner}
\end{figure}

Another interesting related object is a 3-line corner, composed of three mutually intersecting line defects in $\mathbb{R}^{d\geq3}$ (see Fig. \ref{fig:3lcorner}).\footnote{Corners/star-shaped configurations with n-lines were studied in \cite{Henkel1989} for $d=2$.} It stands both on its own and as an emergent object that can appear at the edges of trihedral corners. As before, the action is,\footnote{In this case, to avoid proliferation of indices we assume just a single deformation on each line. Generalization to multiple deformations is straightforward.}
\begin{equation}
	S=S_{\text{CFT}}+\overset{3}{\underset{a=1}{\sum}}h_{a}\int_{\mathcal{D}_{a}}dx\,\mathcal{O}_{a}
\end{equation}
Where we assume that $\Delta_{a}=1$, only for performing the computation. The three line defects are pointing at the directions defined by the unit vectors,
\begin{equation} \label{3l param}
	\mathcal{D}_{1}\left(x\right)=x\hat{e}_{1},\;\mathcal{D}_{2}\left(y\right)=y\hat{e}_{2},\;\mathcal{D}_{3}\left(z\right)=z\hat{e}_{3}.
\end{equation}
\\
There will be several terms appearing in the anomaly coefficient,
\begin{equation} \label{3l-gamma}
	\Gamma=\underset{a<b}{\overset{3}{\sum}}\Gamma_{\text{cusp}}\left(\alpha_{ab}\right)+\Gamma_{\text{3-line}}\left(\alpha_{12},\alpha_{23},\alpha_{13}\right)
\end{equation}
The well-known cusp terms arise at the level of the second order/two-point function in the free energy and are given by \cite{Cuomo:2024psk},
\begin{equation}
	\Gamma_{\text{cusp}}\left(\alpha_{ab}\right)=-\frac{\pi-\alpha_{ab}}{\sin\left(\alpha_{ab}\right)}h_{a}h_{b}+\frac{h_{a}^{2}+h_{b}^{2}}{2}
\end{equation}
In case the cusp is extended (2 intersecting infinite lines), one can obtain the anomalous dimension by direct calculation such as in \eqref{OPE2} or alternatively by gluing together 4 cusps with supplementary angles $\alpha$ and $\pi-\alpha$,
\begin{equation} \label{CAD 2 int}
	\Gamma_{\text{2-lines}}\left(\alpha_{ab}\right)=-\frac{2\pi}{\sin\left(\alpha_{ab}\right)}h_{a}h_{b}
\end{equation}
The focus of this subsection is $\Gamma_{\text{3-line}}$, arising from all three lines meeting mutually at a single point. There are two cases that can be studied, as depicted in Fig. \ref{fig:3lcorner}. For the 3-line corner, we may view each line as the world-line of a point-like impurity via conformal mapping to the cylinder \cite{Henkel1989,Cuomo:2024psk}. In this interpretation one can think of $\Gamma_{\text{3-line}}$ as the three-body potential of impurities living on $S^{d-1}$. We shall derive this potential for the two cases by computing the following term in the free energy,
\begin{align}
	\log Z&=-\Gamma_{\text{3-line}}\left(\alpha_{12},\alpha_{23},\alpha_{13}\right)\log\left(\frac{L}{a}\right)+\dots \notag \\
	&=-h_{1}h_{2}h_{3}\int_{\mathcal{D}_{1}}dx\int_{\mathcal{D}_{2}}dy\int_{\mathcal{D}_{3}}dz\left\langle \mathcal{O}_{1}\left(x\right)\mathcal{O}_{2}\left(y\right)\mathcal{O}_{3}\left(z\right)\right\rangle +\dots
\end{align}
For that task we have to address the following integral,
\begin{equation} \label{3l integral}
	\iiintop\frac{dx\,dy\,dz}{\sqrt{\left(x^{2}+y^{2}-2xy\cos\alpha_{12}\right)}\sqrt{\left(x^{2}+z^{2}-2xz\cos\alpha_{13}\right)}\sqrt{\left(y^{2}+z^{2}-2yz\cos\alpha_{23}\right)}}
\end{equation}
Where the range of integration for the non-extended 3-line corner is over the domain $x,y,z\geq0$. By moving to cylindrical coordinates $\left(z,\rho,\theta\right)$ and rescaling $z$ to a dimensionless variable $z=t\rho$, we rewrite the integral \eqref{3l integral} as $\mathcal{J}\left(\alpha_{12},\alpha_{23},\alpha_{13}\right)\int\frac{d\rho}{\rho}$, where,
\begin{gather} \label{ellipticpre}
	\mathcal{J}_{\frac{\pi}{2}}\left(\alpha_{12},\alpha_{23},\alpha_{13}\right)=\underset{0}{\overset{\pi/2}{\int}}d\theta\frac{I_{\alpha_{13},\alpha_{23}}\left(\theta\right)}{\sqrt{\left(1-\cos\alpha_{12}\sin2\theta\right)}} \\
	\mathcal{J}_{2\pi}\left(\alpha_{12},\alpha_{23},\alpha_{13}\right)=\underset{0}{\overset{2\pi}{\int}}d\theta\frac{I_{\alpha_{13},\alpha_{23}}\left(\theta\right)+I_{\pi-\alpha_{13},\pi-\alpha_{23}}\left(\theta\right)}{\sqrt{\left(1-\cos\alpha_{12}\sin2\theta\right)}}
\end{gather}
And $I_{\alpha,\beta}\left(\theta\right)$ is the following elliptic integral,
\begin{equation} \label{EllipticFormula}
	I_{\alpha,\beta}\left(\theta\right)=\underset{0}{\overset{\infty}{\int}}\frac{dt}{\sqrt{\left(\left(t-\cos\alpha\cos\theta\right)^{2}+\sin^{2}\alpha\cos^{2}\theta\right)\left(\left(t-\cos\beta\sin\theta\right)^{2}+\sin^{2}\beta\sin^{2}\theta\right)}}
\end{equation}
$\mathcal{J}_{\frac{\pi}{2}}$ is used for the 3-line corner, and $\mathcal{J}_{2\pi}$ for the extended one. The integral \eqref{EllipticFormula} has a closed form solution that is fully described in \eqref{EllipticFull}. In conclusion,
\begin{gather}
    \Gamma_{\text{3-line}}\left(\alpha_{12},\alpha_{23},\alpha_{13}\right)=Ch_{1}h_{2}h_{3}\mathcal{J}_{\frac{\pi}{2}}\left(\alpha_{12},\alpha_{23},\alpha_{13}\right) \\
	\Gamma_{\text{ext.-3-line}}\left(\alpha_{12},\alpha_{23},\alpha_{13}\right)=Ch_{1}h_{2}h_{3}\mathcal{J}_{2\pi}\left(\alpha_{12},\alpha_{23},\alpha_{13}\right)
\end{gather}
Where $C$ is the OPE coefficient. For the purpose of demonstration, it is enough to consider the case, ${\alpha_{23}=\alpha_{13}=\frac{\pi}{2},\;\alpha_{12}=\alpha}$, because \eqref{EllipticFormula} simplifies considerably,\footnote{The choice of angles is only a matter of convenience. The result is symmetric in $\alpha_{12},\alpha_{23},\alpha_{13}$, as clearly visible in \eqref{3l integral}.}
\begin{equation}
	I_{\frac{\pi}{2},\frac{\pi}{2}}\left(\theta\right)=\frac{K\left(1-\cot^{2}\theta\right)}{\left|\sin\theta\right|}
\end{equation}
Where $K$ is the Complete Elliptic Integral of the First Kind \eqref{EllipticDef}.
\begin{figure}[h!]
	\centering
	\includegraphics[width=0.45\columnwidth]{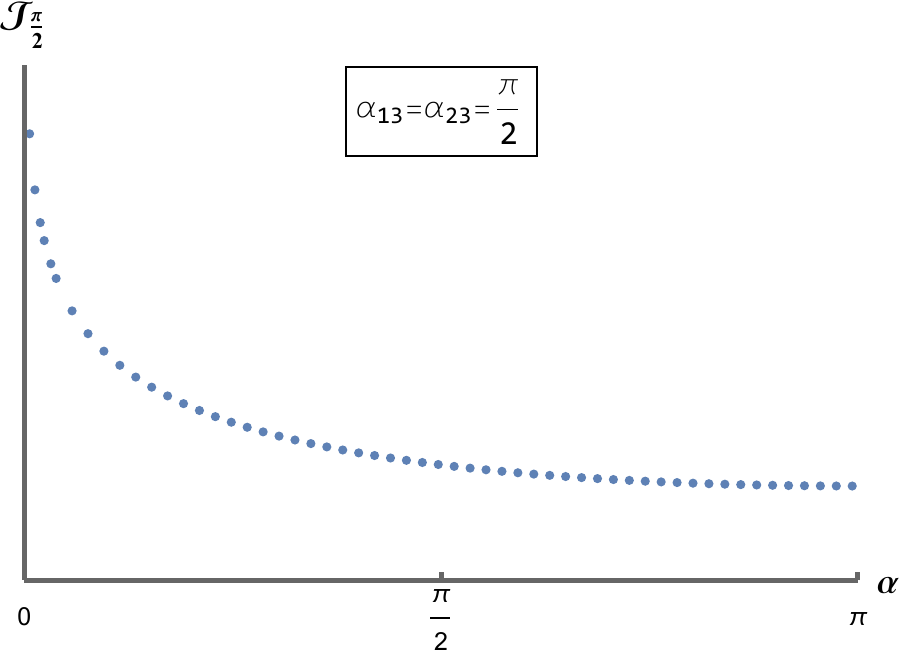}\quad\quad \includegraphics[width=0.45\columnwidth]{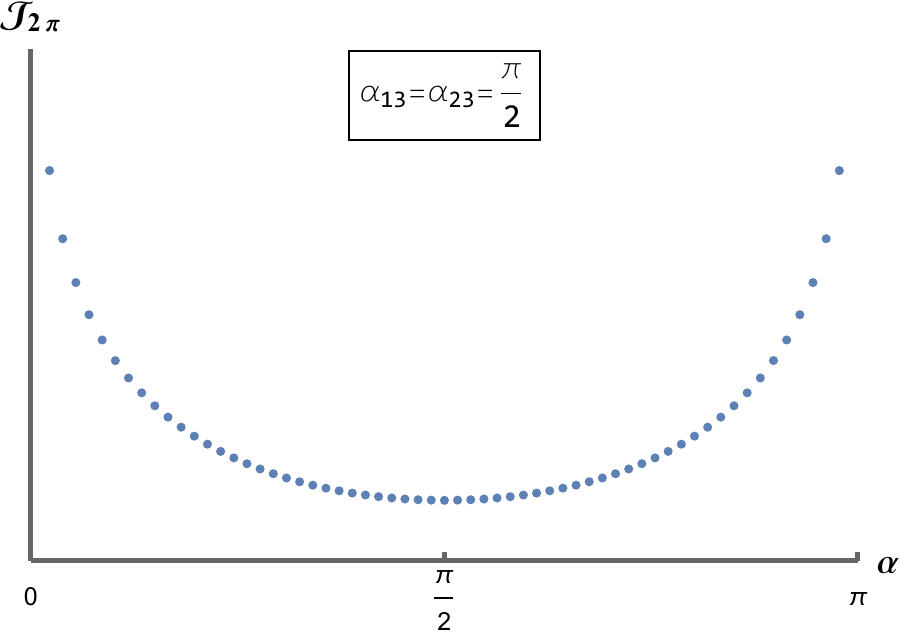}
	\caption{A configuration of 3 line defects with two right angles and $\alpha_{12}=\alpha$ (arbitrarily chosen). \textit{Left Panel}: $\mathcal{J}_{\frac{\pi}{2}}\left(\alpha\right)$ Plotted for the 3-line corner. \textit{Right Panel}: $\mathcal{J}_{2\pi}\left(\alpha\right)$ Plotted for the extended 3-line corner.}
	\label{fig:Elliptic}
\end{figure} \newline
$\mathcal{J}$ can be computed numerically, as demonstrated in Fig. \ref{fig:Elliptic}. We observe that ${\mathcal{J}_{\frac{\pi}{2}}\left(\alpha\rightarrow\pi\right)}$ approaches a non-vanishing value, which is also a minimum due to the reflection symmetry, $\mathcal{J}_{\frac{\pi}{2}}\left(\alpha\right)=\mathcal{J}_{\frac{\pi}{2}}\left(2\pi-\alpha\right)$. This value is related to the non-vanishing Casimir energy of the "T"-shape configuration the 3 line defects assume at  $\alpha=\pi$. For the extended corner, the $\alpha\rightarrow\pi-\alpha$ symmetry is clearly viewable. Finally, notice that as $\alpha\rightarrow 0$ the 3-line potential has the form of $\Gamma_{\text{3-line}}\left(\alpha\right)\approx c_1 \log\alpha+c_2$. This fact is consistent with the notion of fusions of line defects \cite{Cuomo:2024psk,Diatlyk:2024zkk}, as when $\alpha_{12}\rightarrow 0$, the full potential \eqref{3l-gamma} should relate to a single cusp that is formed by $\mathcal{D}_3$ and the fusion product of $\mathcal{D}_1$ and $\mathcal{D}_2$. The $\log\alpha$ term is associated with a slightly irrelevant operator supported on the fused line.
\section{Discussion}
In the $d=3-\epsilon$ tricritical model, we have seen how the operator $\phi^2$ is turned on perturbatively at the edges of the $\phi^4$ defects. We computed the anomalous dimension of the edge operator for a variety of cases, finding a non-trivial dependence on the angle of intersection $\alpha$. It is important to understand how the interactions turned on at the edge affect physical observables such as critical exponents. Critical exponents related to the bulk propagator typically display an angular dependence that is linear in $\frac{\pi}{\alpha}$ to leading order \cite{Cardy1983,Barber1984,Guttmann1984,Cardy1984,Bariev1986,Larsson1986,Saxena1987,Wang1990,Pleimling,PhysRevB.59.65,Pleimling2002,Igloi1993,Bissi:2022bgu}. They were computed primarily in theories where the wedge sets as the boundary of the bulk, usually specified by Dirichlet and or Neumann conditions. As a first step in understanding critical exponents in systems with intersecting defects, it will be wise to study the $O\left(N\right)$ model in $d=4-\epsilon$ with wedges composed of localized $\phi^2$ interactions \cite{10.21468/SciPostPhys.15.6.240,Shachar2024,Trepanier2023a,RavivMoshe2023a,Giombi2023a}. I plan to pursue this direction in the future.

It will also be interesting to expand the findings of this paper to deformations involving fermionic and higher-spin primary operators, especially for the 3-line corner that can also be constructed with vector primaries. There is much more left to understand about both the 3-line and trihedral corners. For start, it is intriguing that the cusp anomalous dimension and the 3-line potential have opposite signs. There is hence a very rich structure for the total potential left to be explored for more concrete models. A further curious observation is that the corner anomaly of 2 intersecting lines \eqref{CAD 2 int} (or extended cusp) and of extended trihedral corners \eqref{CAD-result} have a similar mathematical structure. The structure is clearly related to the inverse volume of the unit Parallelotope subtended by the corner. It will be interesting to understand if this pattern persists for higher-dimensional corners involving more than 3 defects, particularly because in that case the anomaly is already related to four-point functions and higher of the bulk CFT.
\section*{Acknowledgments} 
I would like to thank Michael Smolkin, Zohar Komargodski, Ritam Sinha, Gabriel Cuomo, and Barak Kol for the helpful discussions and comments. The work of T.S. is supported by the Israeli Science Foundation Center of Excellence (grant No. 2289/18).

\appendix
\section{List of Integrals} \label{appA}
Formulas \eqref{form1},\eqref{Feynman} appear in most QFT textbooks (see \textit{e.g.} \cite{Srednicki:2007qs}). Formula \eqref{EllipticFull} is taken from \cite{Byrd1971HandbookOE}.
\begin{equation} \label{form1}
	\text{\ensuremath{\int d^{d}x\frac{\left(x^{2}\right)^{a}}{\left(x^{2}+D\right)^{b}}}}=\pi^{\frac{d}{2}}\frac{\Gamma\left(b-a-\frac{d}{2}\right)\Gamma\left(a+\frac{d}{2}\right)}{\Gamma\left(b\right)\Gamma\left(\frac{d}{2}\right)}\frac{1}{D^{b-a-\frac{d}{2}}}
\end{equation}
\begin{equation} \label{Feynman}
    \frac{1}{A_{1}^{\alpha_{1}}A_{2}^{\alpha_{2}}A_{2}^{\alpha_{2}}}=\frac{\Gamma\left(\alpha_{1}+\alpha_{2}+\alpha_{3}\right)}{\Gamma\left(\alpha_{1}\right)\Gamma\left(\alpha_{2}\right)\Gamma\left(\alpha_{3}\right)}\underset{0}{\overset{1}{\int}}du\underset{0}{\overset{1}{\int}}dv\frac{u^{\alpha_{1}-1}\left(1-u\right)^{\alpha_{2}-1}v^{\alpha_{3}-1}\left(1-v\right)^{\alpha_{1}+\alpha_{2}-1}}{\left[\left(1-v\right)\left(uA_{1}+\left(1-u\right)A_{2}\right)+vA_{3}\right]^{\alpha_{1}+\alpha_{2}+\alpha_{3}}}
\end{equation}
\begin{equation} \label{EllipticFull}
	I\left(\gamma\right)=\underset{\gamma_{1}}{\overset{\gamma}{\int}}\frac{dt}{\sqrt{\left(\left(t-b_{1}\right)^{2}+a_{1}^{2}\right)\left(\left(t-b_{2}\right)^{2}+a_{2}^{2}\right)}}=gF\left(\varphi,k\right)
\end{equation}
Where in \eqref{EllipticFull}, $g$ and $\gamma_1$ are parameters and $F\left(\varphi,k\right)$ is the Incomplete Elliptic Integral of the First Kind with modular angle $\varphi$ and parameter $k$ (the complete elliptic integral is defined by $K\left(k\right)=F\left(\frac{\pi}{2},k\right)$). All are given by,
\begin{gather} \label{EllipticDef}
	F\left(\varphi,k\right)=\underset{0}{\overset{\varphi}{\int}}\frac{d\theta}{\sqrt{1-k^{2}\sin^{2}\theta}} \\
	A=\sqrt{\left(b_{1}-b_{2}\right)^{2}+\left(a_{1}+a_{2}\right)^{2}},\quad B=\sqrt{\left(b_{1}-b_{2}\right)^{2}+\left(a_{1}-a_{2}\right)^{2}} \notag \\
	k^{2}=\frac{4AB}{\left(A+B\right)^{2}},\quad g=\frac{2}{A+B} \notag \\
	g_{1}^{2}=\frac{4a_{1}^{2}-\left(A-B\right)^{2}}{\left(A+B\right)^{2}-4a_{1}^{2}},\quad\gamma_{1}=b_{1}-a_{1}g_{1},\quad\tan\varphi=\frac{\gamma-\gamma_{1}}{a_{1}+g_{1}b_{1}-g_{1}\gamma} \notag
\end{gather}
Notice that there is a discontinuity when the denominator of $\tan\varphi$ vanishes, and the modular angle jumps from $\frac{\pi}{2}$ to $-\frac{\pi}{2}$. The solution \eqref{EllipticFull} remains valid after $\gamma$ crosses that point if one adds the discontinuity by hand to the integral. For example,
\begin{equation}
	\underset{0}{\overset{\infty}{\int}}\frac{dt}{\sqrt{\left(\left(t-b_{1}\right)^{2}+a_{1}^{2}\right)\left(\left(t-b_{2}\right)^{2}+a_{2}^{2}\right)}}=gF\left(\varphi\left(\gamma\rightarrow\infty\right),k\right)-gF\left(\varphi\left(0\right),k\right)+2gK\left(k\right)
\end{equation}

\bibliography{ref}
\bibliographystyle{utphys.bst}

\end{document}